# Single atom force measurements: mapping potential energy landscapes via electron beam induced single atom dynamics

O. Dyck,[1,2] F. Bao,[3] M. Ziatdinov,[1,2] A. Yousefzadi Nobakht,[4] S. Shin,[4] K. Law,[5,6] A. Maksov,[1,2,7] B.G. Sumpter,[1,2] R. Archibald,[1,5] S. Jesse,[1,2] and S.V. Kalinin[1,2]

[1] The Institute for Functional Imaging of Materials, Oak Ridge National Laboratory, Oak Ridge, TN 37831

[2] The Center for Nanophase Materials Sciences, Oak Ridge National Laboratory, Oak Ridge, TN 37831

[3] Department of Mathematics, The University of Tennessee at Chattanooga, Chattanooga, TN, 37403

[4] Department of Mechanical, Aerospace, and Biomedical Engineering, The University of Tennessee, Knoxville, TN 37996

[5] Computer Science and Mathematics Division, Oak Ridge National Laboratory, Oak Ridge, TN 37831

[6] School of Mathematics, University of Manchester, Manchester, UK

[7] Bredesen Center for Interdisciplinary Research and Education, The University of Tennessee, Knoxville, TN 37996

**In the last decade, the atomically focused beam of a scanning transmission electron microscope (STEM) was shown to induce a broad set of transformations of material structure, open pathways for probing atomic-scale reactions and atom-by-atom matter assembly. However, the mechanisms of beam-induced transformations remain largely unknown, due to an extreme mismatch between the energy and time scales of electron passage through solids and atomic and molecular motion. Here, we demonstrate that a single dopant Si atom in the graphene lattice can be used as an atomic scale force sensor, providing information on the random force exerted by the beam on chemically-relevant time scales. Using stochastic reconstruction of molecular dynamic simulations, we recover the potential energy landscape of the atom and use it to determine the beam-induced effects in the thermal (i.e. white noise) approximation. We further demonstrate that the moving atom under beam**

excitation can be used to map potential energy along step edges, providing information about atomic-scale potentials in solids. These studies open the pathway for quantitative studies of beam-induced atomic dynamics, elementary mechanisms of solid-state transformations, and predictive atom-by-atom fabrication.

Controlling and assembling matter on the atomic scale remained one of the central foci of modern sciences, dating back to the renowned talk by Richard Feynman in 1960.[1] The scanning tunneling microscopy (STM) based atomic assembly demonstrated by Don Eigler thirty years later[2] was the first realization of this concept, and, in synergy with advanced surface science methods, has opened the pathway for single atom fabrication,[3-5] for the first time enabling fabrication of solid-state qubits, the enabling element of the quantum computer, with atomic precision. In the last few years, the requirements of the quantum computing race and growing need for beyond Moore technologies make atomic based fabrication one of the central targets for basic and applied research, necessitating both optimization of existing and development of new paradigms for manipulation of matter atom-by-atom.[6-10]

Atomic scale manipulation is inseparable from visualization of matter at the atomic level, and in fact STM was originally introduced as a purely imaging tool. The alternative pathway for atomically resolved imaging is the electron beam based (scanning) transmission electron microscopy ((S)TEM). Following the introduction of aberration correction, atomically-resolved imaging has become routine. Notably, the advancement of STEM has resulted in multiple observations of beam induced changes in matter on the atomic level, where beam induced transformations can be visualized before, during, and after the process. For example, many studies of graphene have noted a variety of defect transformations under e-beam irradiation.[11-23] Additionally, beam induced transformations between graphene and foreign atomic species have also been studied,[24-35] as well as beam induced transformations in other material systems.[36-40] Following these observations, it was proposed that the synergy of electron beam control and real time feedback can be used to devise electron beam driven assembly of matter.[41] Recent developments along this line of investigation have begun to establish the physics behind *in situ* beam-induced atomic processes,[32, 42-45] as well as demonstrate various atomic scale manipulation schemes,[6-10, 46-51] explore bonding and atomic scale electronic structure,[34, 35, 52-54] and model theoretical structures which may be within the grasp of these emerging capabilities.[55, 56]

The progress in this field necessitates understanding of fundamental mechanisms of beam-induced changes in solids on the atomic level. Notably, the theory of the elastic and inelastic scattering in STEM is well developed, as driven by the requirements of STEM image simulation[57] and prediction of electron-energy loss spectroscopy (EELS).[58] However, analysis of the beam-induced damage in solids represents a considerably more complicated problem,[59-61] traditionally analyzed on the level of direct elastic scattering on nuclei (knock-on) or two-temperature models.[62] A number of treatments of high energy processes starting from the formation of non-equilibrium hot electron and hole carriers, thermalization of the electronic subsystem, and energy transfer to the ionic subsystem are also available. However, the extreme mismatch between the time and energy scales of the electron beam (40-300 keV, and ~ attoseconds for electron residence time in the atomic volume) and chemical transformations (~1 eV) and atomic models render the problem extremely complex for forward prediction.

Here, we demonstrate that the observation of electron beam-induced displacements of a single Si impurity atom confined at a defined lattice site in a 2D graphene lattice can be used as a sensitive probe of the energy transfer from the electron beam to the lattice. In this matter, we utilize the impurity as a single-atom force sensor. We develop a stochastic reconstruction method that allows extraction of the free energy landscape from molecular dynamics (MD) simulation of the system and, with known energy landscape, determination of the excitation exerted by the beam in the thermal (i.e. assuming that force is a white noise-like) approximation. We further demonstrate that the observations of the moving atoms allow reconstruction of the free energy landscape along step edges.

Figure 1 shows an example HAADF-STEM image of a single Si substitutional defect in a graphene lattice. It is well-known that the contrast produced by this imaging mode is proportional to the Z number of the atom.[63, 64] The decrease in intensity as the beam is moved away from the atom is due to the diminished intensity of the probe tails. Thus, the atom acts as a delta function convolved with the probe profile. On close inspection of the image, however, we note that the Si atom exhibits a streaky and broken intensity profile. It may be tempting to ascribe this behavior to the (often observed) emission tip noise.[65] Yet, if this were so, such streaky and broken profiles would also be observed on the neighboring C atoms as well. Furthermore, we observed multiple examples of streaky atoms at the adjacent lattice sites (i.e. "dimer" structures), with clear alignment of dark and bright streaks by single lattice translation vector. Lacking another explanation and comparing to similar dynamics often observed in scanning tunneling microscopy (STM) experiments, we conclude that the observation is not microscope instability and results from the Si atom moving slightly toward and away from the beam, which acts to modulate the observed intensity. Depending on exact beam parameters, this beam-induced displacement can be confined to a single lattice site or result in atom jumping between two adjacent lattice sites (which process can give rise to "dimers" or "cut" atoms). Based on this supposition, we can use the line-to-line variations in the image to approximate the dopant atom motion during imaging.

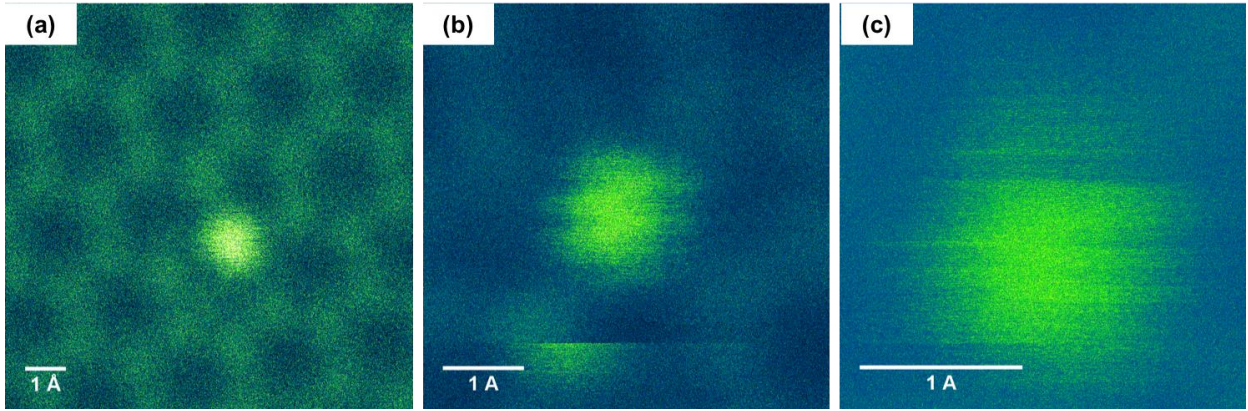

**Figure 1:** (a) HAADF image of a single Si atom in the graphene lattice. (b) Illustration of atomic instability. Note the presence of the "cut" atom in the lower part of the image. (c) Zoom-in illustrating the presence of the clearly visible streaks due to the wobbling motion of the atom during image acquisition. The scan is acquired from left to right (fast scan direction) and from top to bottom (slow scan direction). Images were artificially colored using the "Green Fire Blue" look up table in ImageJ.

The unique aspect of STEM imaging is that the measured image can be, to a good approximation, represented as a convolution of the ideal image with the STEM point spread function.[66] Correspondingly, the local image intensity and position of the maximum along the scan line can be used to infer the information on the position of the atom center with respect to the beam. Here, we analyze a single frame of a raster scanned image of a single dopant atom in a graphene lattice to extract information on the average residence time that the atom spends at particular locations during imaging.

To perform this analysis, it is assumed that the 2D intensity profile of the dopant atom is a radially symmetric Gaussian with a fixed height and width. Next, we fit the intensity profile of each scan line to a noise-floor thresholded 1D Gaussian which has four fitting parameters: amplitude, the central position of the atom in the x-direction ($x_{center}$), width, and noise floor height. An initial fitting is run for each scan line with all four of the parameters free. From this, a max height ($H_{max}$), max width ($w_{max}$), and noise floor height is determined for the whole image. The line-by-line 1D Gaussian fit is run again, this time with the width and noise floor set to the global average. Because we have assumed a radially symmetric 2D Gaussian for the shape of the observed atom, we can combine information about the line-to-line fitted height ($H_{local}$), the overall max amplitude, the atom width, and the scan line position ($y_{local}$) to calculate the line-to-line variation in atom position in the y-direction ($y_{center}$):

$$y_{center} = w_{max}\sqrt{\ln\left(\frac{H_{max}}{H_{local}}\right)} - |y_{local}| \qquad (1)$$

Because the inverse function of the Gaussian is double-valued, some care is needed to make sure that the correct solution for $y_{center}$ is chosen throughout. The final output of this portion of the analysis is an estimate of the central *x,y* position of the atom for each scan line which can be used to yield a map of the likelihood of the atom residing at particular locations.

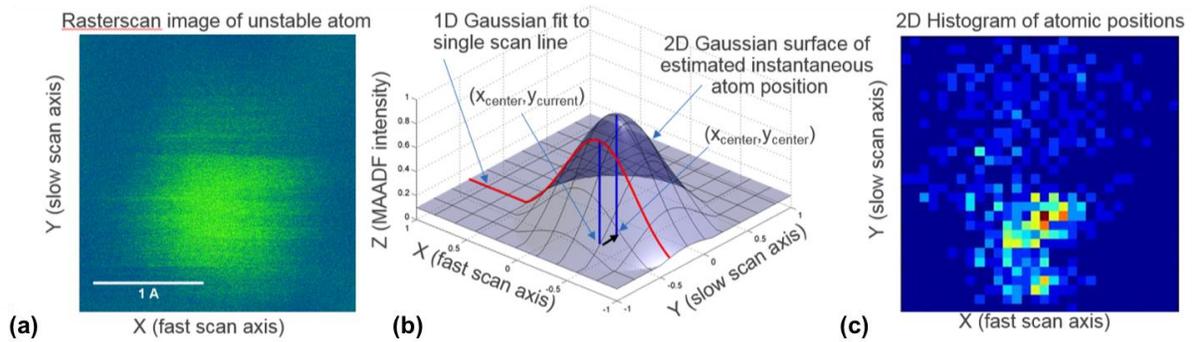

**Figure 2:** Extraction of the observed fluctuations of a single atom in an atomic image. (a) Observed atomic contrast, (b) sketch of the fitting procedure, and (c) derived atomic positions. Note the asymmetry of the probability distribution associated with the scanning motion of the beam.

This analysis yields the information on the atomic position with respect to the beam while the beam scans across the atom, thus defining beam induced dynamics at the ~pm length scales with ~ms time resolution. However, to extract physical meaning and information on the effects induced by the beam it is necessary to combine this data with models of the free energy landscape. In the first approximation, the latter can be estimated as the sum of elastic responses of the individual Si-C bonds, and the force density experienced by the atom can be obtained via elementary arithmetic operation assuming the atomic jumps are uncorrelated.

However, this approach is limited, since the carbon subsystem can also react to the Si atom motion on the time scale of observed vibrations (and below). Furthermore, while STEM data provides the information on the atomic motion in the image plane only, in a realistic material motion in the z-direction is also possible. To avoid these uncertainties, here we adopt a method based on the stochastic reconstruction of the free energy landscape from observed trajectories assuming thermal-like excitation of the atom (see supplemental materials for details [67]). In this approximation, we assume that the atom is excited by white (i.e. having constant spectral density) uncorrelated force, which is equivalent to an effective temperature or deposition of energy, E, active on the atom. This effect can in turn be represented as effective temperature of the atom induced by the beam excitation.

We implement this approach in two steps. As a first step, we perform the molecular dynamic simulation of the Si-C system and reconstruct the free energy landscape from calculated atomic displacements for a ***known*** excitation force. In the second step, we determine the unknown excitation force due to beam-nucleus interaction from the experimentally observed atomic dynamics.

The MD simulations were performed to simulate the Si-C atom response to a known force applied to a single dopant Si atom embedded in a monovacancy in a graphene lattice. For this purpose, a periodic force with an exponential distribution was applied to the Si atom and the potential energy variations were recorded during the simulations (Figure 3). For more details about the MD simulation methodology refer to the supplementary materials. This modeling yields (*x,y,z*) atomic positions as a function of time.

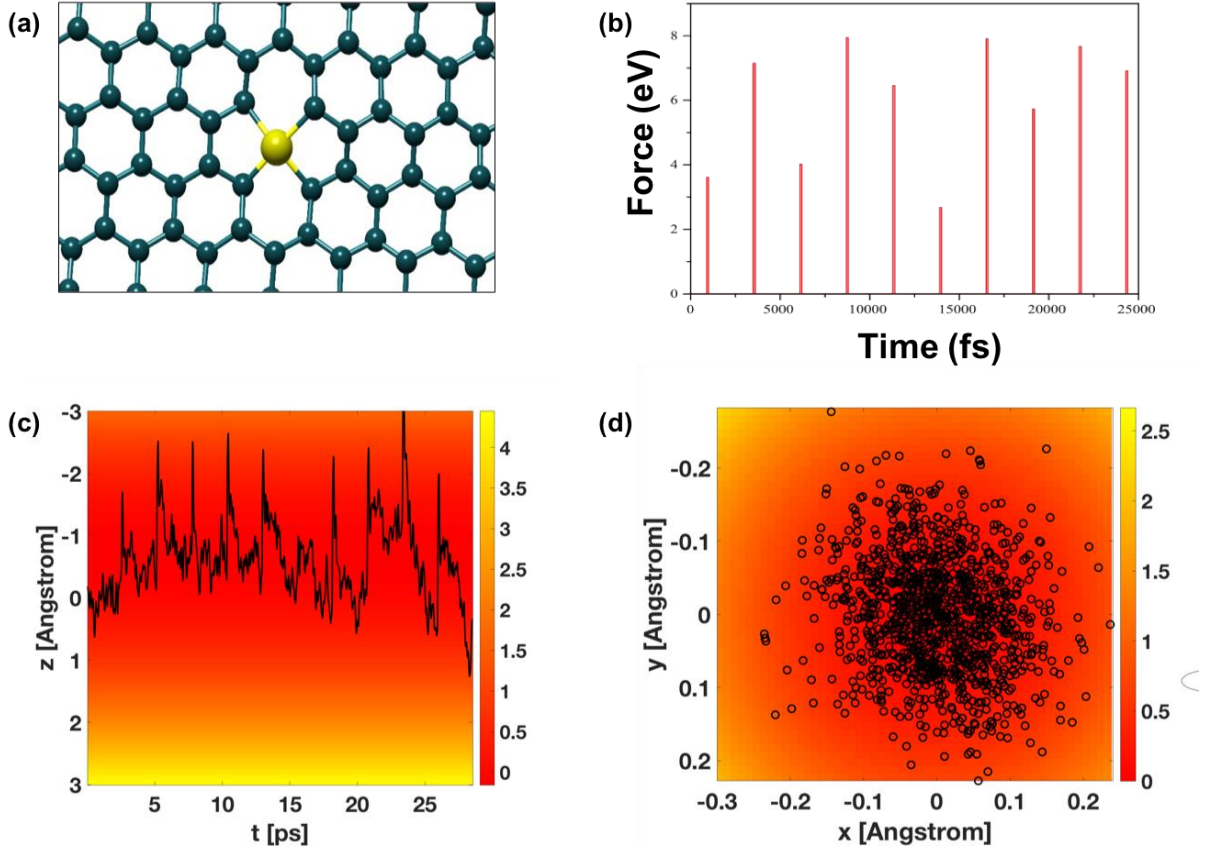

**Figure 3**: a) Silicon atom configuration in graphene sheet. b) Probing force distribution applied to Si atom. c) $e^{-V_z/E_z}$ and trajectory of $(z_t)$. d) $e^{-V_\perp/E_\perp}$ and trajectory of $(x_t, y_t)$. The energy scale (color) in (c,d) is in eV.

To reconstruct the potential over the $(x, y, z)$ space, we choose the model potential consistent with site symmetry as $V(x, y, z; \mathbf{t}) = V_\perp(x, y) + V_z(z)$, where the following parametric ansatz are employed for the individual terms

$$V_\perp(x, y) = a(x^2 + y^2)[1 + b\cos(4\tan^{-1}(x/y))], \quad V_z(z) = cz^2 + dz, \qquad (2a,b)$$

where $\mathbf{t} = (a,b,c,d)$ are parameters. Note that in MD simulation all $(x,y,z)$ coordinates are available during the modeling, unlike the STEM experiment which is sensitive only to $(x,y)$.

For femto-scale dynamics of the MD simulation, on the order of the decorrelation time of the underlying process, it makes sense to utilize dynamical information in the reconstruction. We leverage the sequential nature of the data and the dynamical information by using a sequential Monte Carlo sampling method to process the observed atomic trajectories and perform parameter estimation. First the energy, E, is fit to the quadratic variation of the process. As mentioned above, it is assumed that the potential $V$ has a specific form and is governed by a set of parameters $\mathbf{t}$. It is furthermore assumed that the data arise as exact observations from a discretized first order Langevin stochastic differential equation (SDE) in this potential,

$$\zeta dX_t = -\nabla V(X_t)dt + \sqrt{2\zeta E}dW_t,$$

where $X_t$ is the state, $E = k_B T$ is the characteristic energy in units of [eV], $\zeta$ is a fluctuation-dissipation constant with units of mass/time, and $W_t$ is a standard Brownian motion with units of $[\sqrt{s}]$.[67] This equation has invariant measure $\exp(-V/E)/Z$, where Z is a normalizing constant.

In this context we recover a Bayesian posterior distribution with a simple tractable form, and we can then implement a sequential Monte Carlo sampling framework [68] by leveraging the sequential form of the posterior. We can further reduce the computational expense by introducing a pseudo-dynamics on the parameters in the spirit of [69] and implementing a standard particle filter. The former results are presented for reconstruction of the posterior potential V given the MD simulation data. The observed atom positions from STEM are separated by much longer times than the scale of the MD. Due to decorrelation in the stochastic model above, we therefore assume they comprise independent and identically distributed observations from $\exp(-V/E)/Z$, which allows us to reconstruct E. Based on this analysis, we find E = 0.26 eV → T = 3017 K. Details about the Bayesian parameter estimation of the molecular potential can be found in the supplementary materials.

The characteristic information on the effective parameters of electron beam induced atomic excitation can be used to map the free energy landscape experienced by the atom during more complex dynamic processes, e.g. atomic motion between adjacent lattice sites.

As an example of application of this for analysis of beam-induced transformations, we apply it to reconstruct free energies along a graphene step edge. In this example, a hole in one layer of bilayer graphene was found with two dopant atoms passivating the step edge. The difference in intensity between the two atoms indicates they are of different atomic species, the brighter one having a higher atomic number. A series of images was acquired, exposing the system to the 60 keV electron beam irradiation while simultaneously recording the positions of each atom.

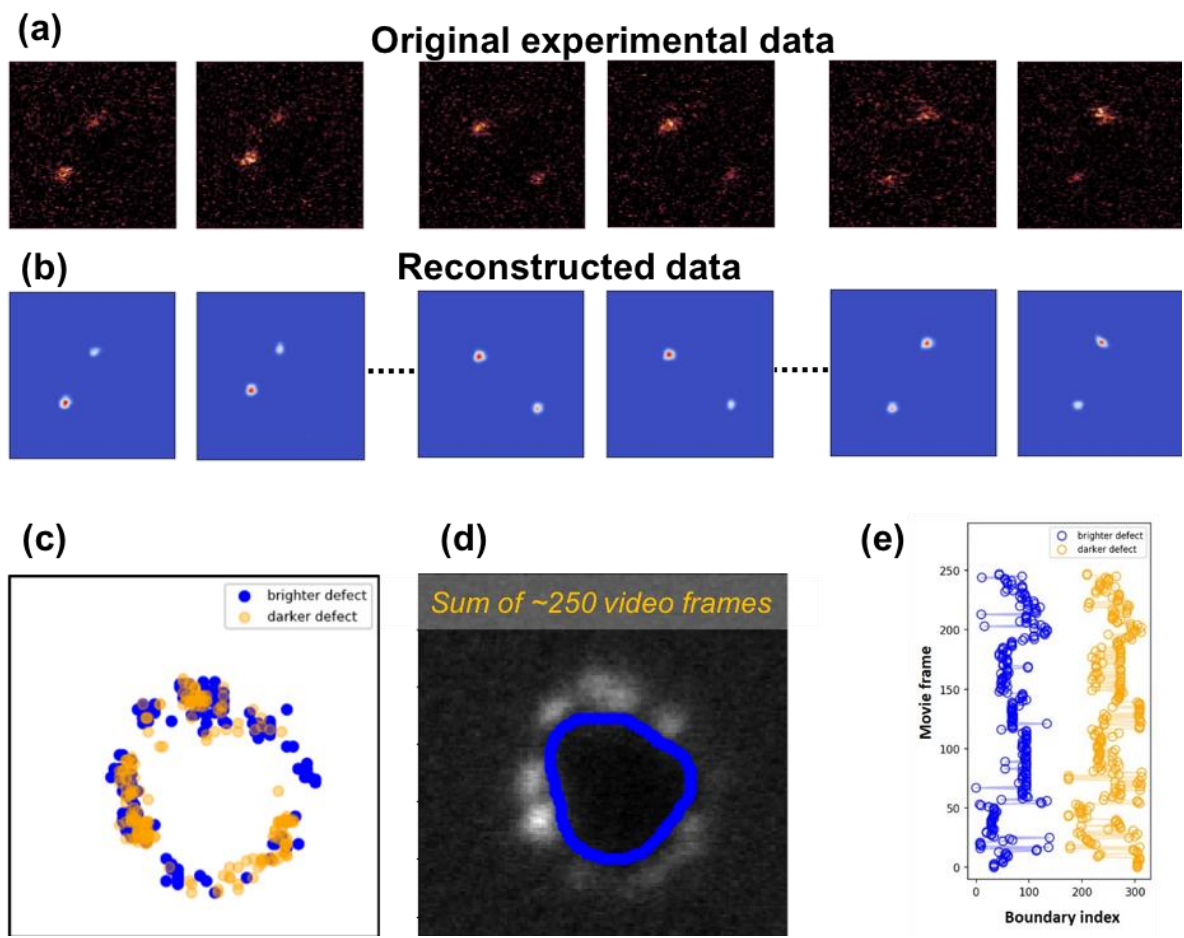

**Figure 4: Deep convolutional autoencoder based analysis of impurity trajectories in graphene.** (a) Representative image frames of the experimental STEM "movie" with two atomic impurities. (b) Output of convolutional de-noising autoencoder for image frames shown in (a). Both encoder and decoder parts of the autoencoder consisted of 3 convolutional layers, with 16 filters of size 3 px by 3 px in each layer, separated by max pooling layer (encoder) and un-pooling layer (decoder). (c) Plot of coordinates of atomic positions for brighter and darker atoms extracted for each of 248 movie frames. (d) Average hole contour overlaid onto the image representing a sum of about 248 movie frames. (e) Analysis of impurity atoms trajectories along the edge of hole represented as a function of the coordinate along the edge (boundary index).

The high level of noise in the experimental series of images (hereafter, "movies") of impurity motion (Fig. 4a) did not allow us to extract positions of the atomic species in each frame using standard image analysis tools (*e.g.*, median filtering and thresholding). To alleviate this issue, we employed a deep convolutional de-noising autoencoder (CDAE) for processing the raw experimental data (Fig. 4b).[70,71] The CDAE is a powerful tool for reconstructing missing/corrupted data from the images, including images where signal and noise can be barely differentiated by the human eye. We used a CDAE model in which both encoder and decoder parts consist of three convolutional layers separated by a max pooling layer (encoder) and an un-pooling layer (decoder). The CDAE was trained on simulated STEM images corrupted with the levels of noise comparable to those observed in the experiment. We then applied the trained CDAE to the real STEM "movies", which allowed us to reconstruct noisy experimental images ("movie" frames) into the images with two well-defined atomic impurities of different intensities ("brighter" and "darker") on a uniform background (Fig. 4b). Thus, with the help of CDAE the problem has been reduced to a trivial problem of tracking a motion of two blobs of different intensities across the frames of the STEM "movie". The extracted positions of atomic impurities from all the "movie" frames are plotted in Fig. 4c. We then extracted an approximate contour of the hole edge (Fig. 4d) assuming the hole's boundary does not undergo any drastic reconstructions during the imaging and plotted an impurity trajectory as a function of coordinate along the edge (boundary index) for both the "brighter" and "darker" atom (Fig. 4e). The impurity dynamics show the character of a telegraph process, in which there are long periods of stability with rapid "jumps" in-between.

As mentioned above, the timescale of microseconds is effectively macroscale with respect to the femto-second dynamics and nanosecond pings, so the observations are considered to be independent and identically distributed (i.i.d.) realizations from the invariant distribution of the underlying femto-scale process. Here we have dynamics of two interacting processes, a dark atom $S_d(t)$ and a bright atom $S_b(t)$, which are assumed to evolve on the torus parametrized by s in [0,1], in units of the circumference of the ring, which is estimated to be between 24 and 28 Angstroms. A non-parametric kernel density estimate is used to reconstruct an approximation of the invariant density $p(s_d, s_b)$=exp(-U($s_d, s_b$))/Z. A Gaussian kernel is used for the non-parametric density reconstruction, with standard deviation l=0.05. Figure 5 shows the marginal reconstructed densities (on each of $s_d$ and $s_b$) after the full 248 observations. The joint density, which describes the correlations between the processes, is presented in the supplementary material.

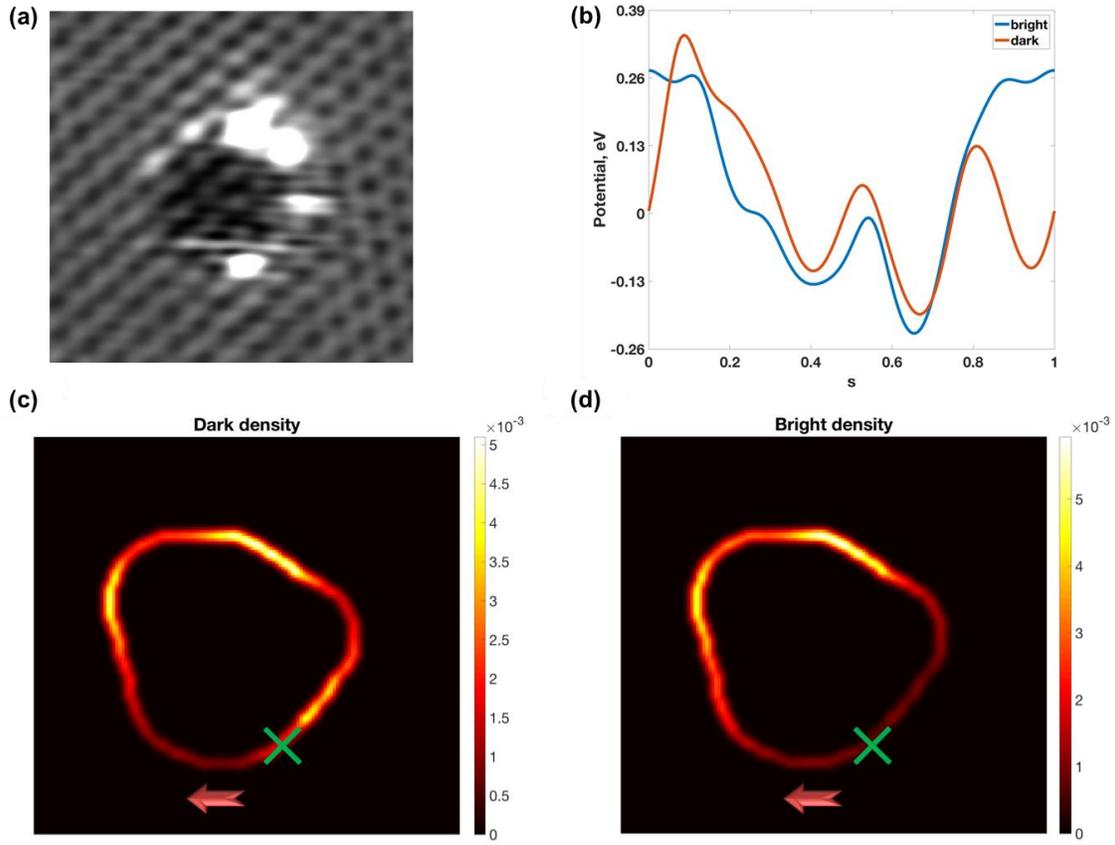

**Figure 5:** (a) Experimental slow-scanned image of the hole prior to the acquisition of the "movie". Fourier filtering was applied to enhance the lattice periodicity. Note the clearly visible structure of bilayer graphene around the hole, atomic structure of graphene in the hole, and "blobs" formed due to the movement of atoms during imaging. (b) Potential energy landscape experienced by the atoms along the hole edge and schematic graphene configuration, (c) reconstructed marginal probability density of the dark atom, d) reconstructed marginal probability density of the bright atom, where p $\propto e^{-U}$. Green crosses mark the origin and arrows show direction of observations.

To summarize, here we demonstrate an approach to reconstruct the force exerted by the electron beam using a single impurity atom with a known free energy landscape as a force sensor. We approximate the electron beam effect on atomic dynamics via a Gaussian process, which is equivalent to an effective temperature approximation, and determine this parameter from the experimental observation of single site atomic dynamics. We further use this approach to map the free energy landscape through which atom travels and relate it to structure.

We note that future opportunities include extending this approach for non-Gaussian models for beam-induced atomic dynamics, where the frequency spectrum of the excitation force has non-trivial frequency dispersion. Furthermore, we envision the determination of the excitation parameters as a function of beam position vs. nucleus, and mapping in the z-direction via a tilt series. Finally, of interest is the evolution of the effective beam parameters describing energy transfer to an atomic nucleus as a function of experimentally controlled parameters such as beam current and energy.

Overall, this study sets the approach for the description of experimentally observed atomic dynamics under the action of an electron beam, and hence is consistent with the description of the multitude of processes reported in the last several years including crystallization of amorphous materials,[72-74] elastic-plastic transition,[75] ferroelectric domain switching,[76] phase transitions,[77, 78] vacancy formation and dynamics,[79] creation and inversion of molecular bonds,[80,81] atomic motion,[82] sculpting,[83] and liquid electrochemistry.[84] This will enable the e-beam to be used as a controllable (or at least understood) probe in the beam induced transformations, providing a new and powerful probe of the atomic world.

**Acknowledgements:** Research was performed at the Center for Nanophase Materials Sciences, which is a US Department of Energy Office of Science User facility. Experimental work was supported by the Laboratory Directed Research and Development Program of Oak Ridge National Laboratory, managed by UT-Battelle, LLC for the U.S. Department of Energy (O.D., S.V.K., S.J.) This research was also sponsored by the Applied Mathematics Division of ASCR, DOE; in particular under the ACUMEN project (F. B., K. J. H. L., R. A.). A.M. acknowledges fellowship support from the UT/ORNL Bredesen Center for Interdisciplinary Research and Graduate Education.

**Supplemental materials**

1. **MD Simulations**

Periodic boundary conditions are applied in the all directions. To avoid interactions between periodic cells in *z* direction, a large vacuum (50 nm) considered in both sides of graphene sheet in this direction. Here, *x* and *y* are in-plane directions and *z* is out of plane direction. A graphene sheet with size of 12×12 nm was used and single Si atom in a monovacancy is considered in the graphene sample. (Fig. S1). Adaptive intermolecular reactive empirical bond order (AIREBO)[1] potential is used to model C-C bonded interactions, and the Tersoff potential[2] was employed for modeling Si-C interactions. Atomic/Molecular Massively Parallel Simulator (LAMMPS)[3] was used to carry out the molecular dynamics simulations; the total simulation time and time step size for all simulations are chosen to be 5.0 ns and 0.5 fs, respectively.

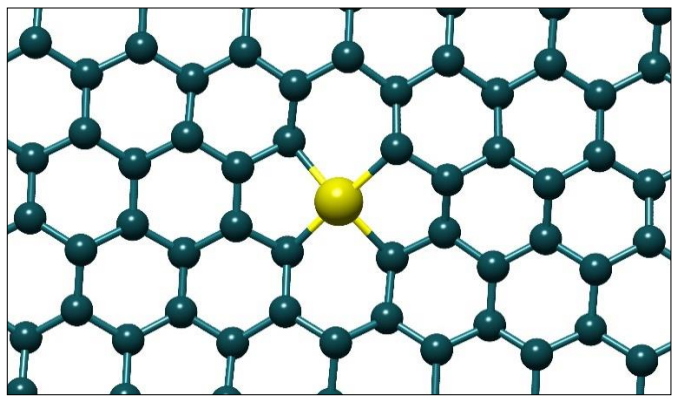

**Figure S1.** Silicon atom configuration in graphene sheet.

The initial structure was allowed to relax at the beginning of the simulation using an NPT (isothermal-isobaric) ensemble for 0.5 ns, which was followed by a NVT ensemble at a temperature of 300K for another 0.5 ns to achieve steady temperature (Temperature fluctuations < 3K) for entire structure. After reaching a steady temperature for the system, a periodic force is applied, with the form

$$F = A(e^{0.5 \times m^2})$$

was applied. Here *F* is the effective force applied to the Si atom in –*z* direction and *m* is a random generated number and *A* is the factor defining the force magnitude. The force was applied as a short delta function with a duration of 100 fs and at a rate of every 2.5 ps. The time gap between each force application is long enough so the effect of the previous force is damped by the system. Fig. S2 shows an example of the applied force distribution.

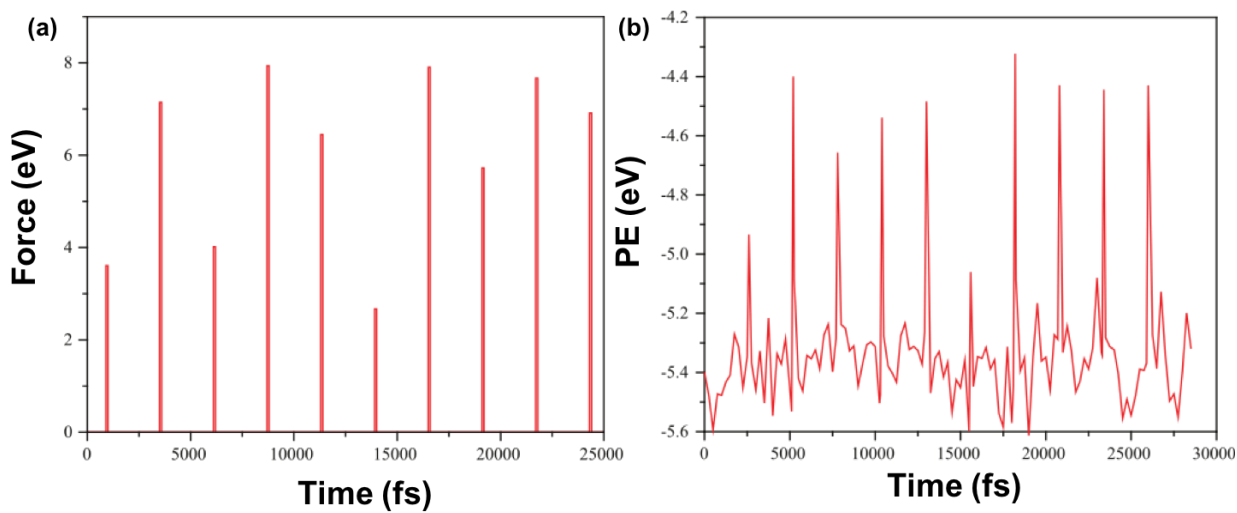

**Figure S2**: a) Force distribution applied to Si atom. b) Potential energy of Si atom with respect to time.

In order to calculate potential, the position and potential energy of the Si atom was exported every time step during the NVE ensemble. The potential energy variation with respect to time for the Si atom under the influence of the force shown in Figure S2 a) is illustrated in Figure S2 b).

2. **Bayesian parameter estimations of the molecular potential**

A somewhat non-standard and ad hoc statistical model is introduced which simplifies the problem of static parameter estimation for stochastic differential equations (SDEs). In particular, discretization error is ignored and observations are assumed exact, which yields a tractable posterior distribution. A sequential Monte Carlo sampler is then implemented for inference from the proposed posterior distribution.

*a: Setup*
Consider the following statistical model

$$\gamma_n(d\theta) = \prod_{i=1}^{n} p_\delta(x_i|x_{i-1},\theta)p_0(d\theta),$$

where $p_0$ is a prior on some parameter $\theta \in \mathbb{R}^d$, and $p_\delta$ is a $\delta$ time-step Euler approximation of the following SDE

$$dX_t = -\nabla U(X_t;\theta)dt + \sqrt{2D}dW_t, \qquad (1)$$

where $D \in \mathbb{R}$ is a diffusion constant, $W_t$ is Brownian motion on $\mathbb{R}^p$, $X_t \in \mathbb{R}^p$, and $a:\mathbb{R}^{p+d} \to \mathbb{R}^p$ is Lipschitz. That is

$$p_\delta(x_i|x_{i-1},\theta) = \exp(-\frac{1}{4D\delta}|x_i - x_{i-1} + \nabla U(X_t;\theta)\delta|^2). \qquad (2)$$

Define the posterior density of $\theta|x_0, \ldots, x_n$ by

$$\eta_n(d\theta) = \gamma_n(d\theta)/\gamma_n(1).$$

The objective is to sample $\theta_i \sim \eta_J$ for $i = 1, \ldots, N$, and approximate expectations for bounded functions $\varphi: \mathbb{R}^d \to \mathbb{R}$

$$\eta_n(\varphi) := \int_{\mathbb{R}^d} \varphi(\theta)\eta_n(d\theta) = \int_{\mathbb{R}^d} \varphi(\theta)\eta_n(\theta)d\theta,$$

by

$$\eta_n(\varphi) \approx \frac{1}{N}\sum_{i=1}^{N}\varphi(\theta^i).$$

We cannot sample from this distribution but we can obtain a convergent estimator using SMC samplers,[4] as described below.

*b: Inferring $\theta$*

Suppose D is known for now. It is well-known how to estimate D for this type of model, given $(x_0, \ldots, x_J)$, which will be done ahead of time (see Section c).

Define

$$G_n(\theta) = \gamma_{n+1}(\theta)/\gamma_n(\theta) = p_\delta(x_n + 1|x_n, \theta).$$

Let $M_n$ denote an MCMC kernel such that

$$(\eta_n M_n)(d\theta) := \int_{\mathbb{R}^d} \eta_n(d\theta')M_n(\theta', d\theta) = \eta_n(d\theta).$$

Let $\theta_0^{(i)} \sim p_0$. For $n = 1, \ldots, J$, repeat the following steps for $i = 1, \ldots, N$:

- Define $\widetilde{w}_n^i := G_{n-1}(\theta_{n-1}^{(i)})$ and $w_n^i = \widetilde{w}_n^i / \sum_{j=1}^N \widetilde{w}_n^i$.
- Resample. e.g. select $I_n^i \sim \{w_n^1, \dots, w_n^N\}$, and let $\widehat{\theta}_n^{(i)} = \theta_{n-1}^{I_n^i}$.
- Mutate. Draw $\theta_n^{(i)} \sim M_n(\widehat{\theta}_n^{(i)}, \cdot)$.

Define

$$\eta_n^N(\varphi) := \frac{1}{N} \sum_{i=1}^N \varphi(\theta_n^{(i)}).$$

Under mild conditions it is well known that as $N \to \infty$, one has that $\eta_n^N(\varphi) \to \eta_n(\varphi)$ almost surely. Rates, central limit theorem, and large deviations estimates can also be obtained.[5]

*c: Inferring D*

Observe that $x_{n+1} - x_n \sim N(-\delta \nabla U(X_t; \theta), 2D\delta)$. It is clear then that the drift term is higher order in $\delta$ when computing an approximation of the quadratic variation of the limiting SDE

$$\widehat{Q} := \frac{1}{J+1} \sum_{n=0}^J (x_{n+1} - x_n)^2.$$

Indeed, this gives a very good approximation to $2D\delta$, and we define $\widehat{D} := \widehat{Q}/(2\delta)$. As $\delta \to 0$, one has that $\widehat{D} \to D$. With a known temperature, i.e. value of $k_b T = E = \zeta D$, we will then be able to estimate the fluctuation-dissipation coefficient $\zeta$ by $\widehat{\zeta} = E/\widehat{D}$.

*d: Filtering with pseudo-dynamics*

An alternative approach to parameter inference is to introduce a pseudo-dynamic on the parameter as follows

$$\theta_{n+1} \sim N(\theta_n, C_\theta),$$

and then solve the filtering problem for $\theta_n | x_1, \dots, x_n$,[6] using a standard particle filter. We find that this approach can provide a reasonable approximate reconstruction with appropriately chosen $C_\theta \propto \delta$.

*e: Inference at the "macroscale"*

Here we consider data points separated by 100s or 1000s of decorrelation times. Note that under appropriate assumptions this implies $X_t \to \rho$ in distribution, where

$$\rho(x;\theta) = \frac{1}{Z}\exp(-U(x;\theta)/D). \qquad (3)$$

Indeed, one has convergence of ergodic averages

$$\frac{1}{T}\int_0^T \varphi(X_t)dt \to \mathbb{E}_\rho(\varphi). \qquad (4)$$

In this case, the same methodology is applicable, except equation (2) will not make sense. It would be very challenging to compute the transition density with a reasonable time-step, as it would require marginalization over all the intermediate steps. However, in light of equation (4), in this case it is reasonable to assume the data points are independent and identically distributed (i.i.d.) and to replace equation (2) with

$$p(x_i|x_{i-1},\theta) = \exp(-\frac{U(x_i,\theta)}{D}).$$

In our case, the macroscale is given ironically by milliseconds, or even micro-seconds, i.e. the scale on which experimental observations are made.

*f: Other approaches*

The other approaches considered here include non-parametric estimation, and a mixed approach consisting of joint non-parametric estimation with nonlinear regression.

The non-parametric estimation proceeds by smoothing an empirical distribution through convolution with a Gaussian kernel

$$K(x;\ell) \propto \exp\{-\frac{1}{2\ell^2}|x|^2\}.$$

In this case, one constructs an empirical measure based on the data points $\{x_i\}_{i=1}^N$ as follows (there is an implicit i.i.d. assumption here)

$$\widehat{\rho}^N := \frac{1}{N}\sum_{i=1}^N \delta_{x_i},$$

and then aims to reconstruct equation (3) by

$$\widehat{\rho}_K(x) := K \star \widehat{\rho}^N = \int K(x-y)\widehat{\rho}^N(y)dy = \frac{1}{N}\sum_{i=1}^N K(x-x_i;\ell).$$

From here, assuming that $K$ is normalized, one has an estimate for $U/D$ as well

$$\widehat{U}_K(x) := -\log \widehat{\rho}_K(x). \tag{5}$$

The second mixed approach proceeds as follows. The estimate in equation (5) can now be fit to a parametric form $U(\cdot;\theta)/D$, using nonlinear regression along a set of points

$$x \in \mathcal{X} := \{x_{min}, x_{min} + \Delta, \ldots, x_{max} - \Delta, x_{max}\}.$$

That is, one minimizes

$$\Phi(\theta) := \sum_{x \in \mathcal{X}} |\widehat{U}_K(x) - \frac{U(x;\theta)}{D} - \log(\int e^{-\frac{U(x;\theta)}{D}} dx)|^2,$$

and you take $\theta^* := \mathrm{argmin}\, \Phi$ as the estimator. The integral is approximated with a quadrature rule. This function could also be taken as a likelihood in a Bayesian framework.

Naturally, in light of the discussion in section 3e, the framework presented in this section makes more sense for the "macroscale" data, i.e. those data which are observed in the experiments. The estimator (5) is used to approximate the effective joint potential for the interacting system of light and dark atoms, from experimental observations $\{x_i\}_{i=1}^N$. In this case $x = (z_1, z_2)$, where $z_i \in [0,1)$, with $i = 1,2$ corresponding to light and dark atom positions, respectively.

### g: *Results*

Now the long MD trajectory is used to fit an effective Langevin dynamics.[7] It is assumed that in this small mass and long time regime it is reasonable to ignore inertial effects and employ a first order Langevin equation for fitting.[7] First the method of section c is employed to predict the diffusion constant from the quadratic variation. It is observed that the dynamics is smooth on a femtosecond timescale, and so a timestep of $\delta = 0.025$ ps is considered. An additional difficulty arises here due to the deterministic time-dependent forcing in the $z$ direction, denoted by $F(t) = (0,0,\mathrm{f}(t))^T$, as well as the difference in effective diffusion coefficients in the $(x,y)$ and $z$ directions, denoted $D_\perp$ and $D_z$, respectively. The Langevin dynamics now take the form

$$dX_t = -(\nabla U(X_t) + F(t))dt + \sqrt{2\boldsymbol{D}}dW_t, \tag{6}$$

where $X = (x,y,z)^T$, and $\boldsymbol{D} = \mathrm{diag}\{D_\perp, D_\perp, D_z\}$. Note that the equations (6) and (1) balance Å. In particular, (1) has been multiplied through by $\boldsymbol{\zeta}^{-1}$, the inverse of the fluctuation-dissipation coefficient, which is now an operator $\boldsymbol{\zeta} = \mathrm{diag}\{\zeta_\perp, \zeta_\perp, \zeta_z\}$. Following a similar procedure as the one described in section c, we can estimate $\zeta_\perp = E/D_\perp$ and $\zeta_z = E/D_z$, by substituting in T=273K, i.e. E = 0.024 eV. In particular, the estimates using the method in section c are $D_\perp = 0.16$ Å$^2$/ps and $D_z = 0.55$ Å$^2$/ps. Since $E = 0.024$ eV, this gives $\zeta_\perp = 0.15$ eV ps/Å$^2$ and $\zeta_z = 0.044$ eV ps/Å$^2$.

Comparing (1) with (6), one can observe furthermore that $\nabla U(X_t) = \zeta^{-1}\nabla V(X_t)$. It is assumed for simplicity that the dynamics obey an invariant measure, despite the time-dependent periodic forcing F(t). It is furthermore assumed that the potential is separable as $U(x, y, z) = U_\perp(x, y) + U_z(z)$, where the following parametric ansatz are employed for the individual terms

$$U_\perp(x, y) = a(x^2 + y^2)\left[1 + b\cos\left(4\tan^{-1}\left(\frac{x}{y}\right)\right)\right], \qquad U_z(z) = cz^2 + dz,$$

and so the invariant measure is proportional to $\exp(-(U_\perp/D_\perp + U_z/D_z))$.

The method of section b is used to fit $(a, b, c, d)$. Note that degeneracy may occur if b is allowed to change sign, and ill-posedness will result from $a, c < 0$. So, it is assumed that $a = e^{\theta_1}$, $b = e^{\theta_2}$, $c = e^{\theta_3}$ and $d = \theta_4$. A prior $\theta \sim N(0, C_0)$ is used, with $C_0 = \sigma^2 I$, and $\sigma = 2$. The parameters $a\zeta_\perp, c\zeta_z$ have units eV/(Å$^2$), $d\zeta_z$ has units of eV/(Å), and b is dimensionless.

The resulting posterior after $n = 1141$ observations separated by $\delta = 0.025$ ps is presented in Figure S3. The posterior expectations on the parameters are given by $\mathbb{E}a = 13.7$, $\mathbb{E}b = 0.0125$, $\mathbb{E}c = 0.882$, and $\mathbb{E}d = 0.479$. It is interesting to observe that $d \approx -\frac{1}{T}\int_0^T f(t)dt$. In the end, the posterior mean potential in eV is given by $V_\perp = U_\perp \zeta_\perp$ and $V_z = U_z \zeta_z$, i.e.

$$V_\perp(x, y) = 2.06(x^2 + y^2)\left[1 + 0.0125\cos\left(4\tan^{-1}\left(\frac{x}{y}\right)\right)\right], \qquad V_z(z) = 0.039z^2 + 0.021z.$$

These are plotted in Figures 3 c) and 3 d) with observation trajectory overlayed.

Now, this potential can be used together with data obtained from observed fluctuations of a single atom as presented in Figure 2 in order to infer the effective temperature. A non-parametric optimization approach similar to the one described in section f is used to fit the curvature at the mode, which yields $E = 0.26$ eV $\to T = 3017$ K. There is 10% relative error in the reconstruction, in L2 norm.

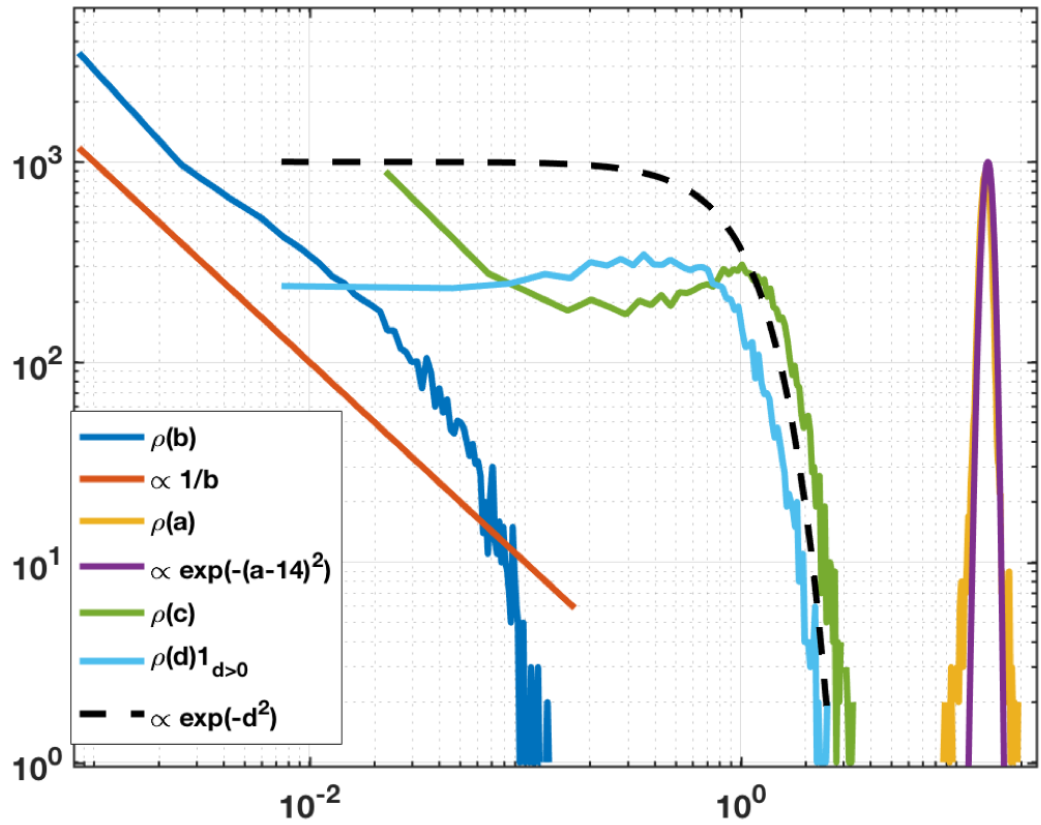

**Figure S3**. Marginal posterior densities of the pushforward of $\boldsymbol{\theta}|\mathbf{x_1}, \ldots, \mathbf{x_n}$ to (a,b,c,d). The results are on a log scale and compared to some functional forms.

The joint potentials arising from the reconstructions of the density $p(s_d, s_b)=\exp(-V(s_d, s_b))/Z$, as described in the main text, after 1, 10, 100, and full 248 observations are shown in figure S4. The resulting potential is dimensionless.

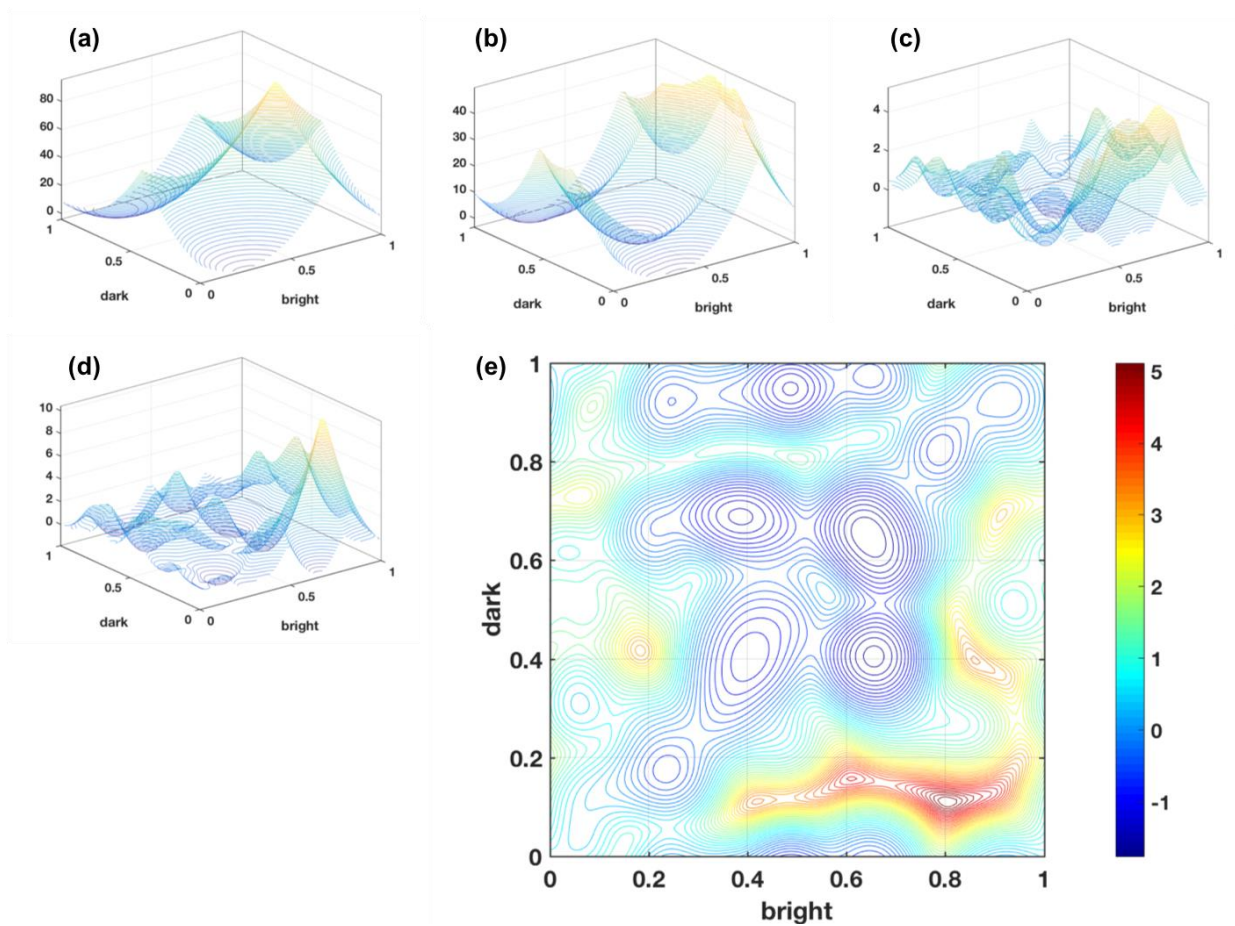

**Figure S4.** Reconstructions of potential on a torus expanded coordinate system a) after 1, b) after 10, c) after 100, and d) after 248 observations. e) Surface plot of the final reconstruction.